\documentclass[12pt,a4paper]{article}
\usepackage{float}
\usepackage{subfig}
\usepackage{amsmath}
\usepackage[dvipsnames]{xcolor}
\usepackage{amssymb}
\usepackage{graphicx}
\usepackage{cancel}
\usepackage{comment}
\usepackage{hyperref}
\usepackage[compat=1.0.0]{tikz-feynman}
\usepackage{csquotes}
\usepackage{multirow}
\usepackage{multirow}
\usepackage{pdflscape}
\usepackage{appendix}
\usepackage{fmtcount} 
\usepackage{array,multirow}

\newcolumntype{C}[1]{>{\centering\arraybackslash}m{#1}}
\usepackage{epsfig}
\usepackage{float}
\usepackage{geometry}
\usepackage{amsmath}
\usepackage{cite}
\usepackage{ctable}
\newenvironment{rcases}
{\left.\begin{aligned}}
{\end{aligned}\right\rbrace}
\newgeometry{vmargin={20mm,20mm},hmargin={20mm,20mm}}
\begin{document}

\title{\textbf{Minimal Type-I Dirac seesaw and Leptogenesis under $A_{4}$ modular invariance}}
\author{Labh Singh$^{1}$\thanks{sainilabh5@gmail.com}, Monal Kashav$^{2}$\thanks{monalkashav@gmail.com}, and Surender Verma$^{1}$\thanks{s\_7verma@hpcu.ac.in}}
\date{%
$^{1}$Department of Physics and Astronomical Science, Central University of Himachal Pradesh, Dharamshala-176215, India\\
$^{2}$Theoretical Physics Division, Physical Research Laboratory, Navarangpura, Ahmedabad-380009, India\\
}
\maketitle
\begin{abstract}
\noindent We present a Dirac mass model based on $A_{4}$ modular symmetry within Type-I seesaw framework. This extension of Standard Model requires three right-handed neutrinos and three heavy Dirac fermions superfields, all singlet under $SU(2)_{L}$ symmetry. The scalar sector is extended by the inclusion of a $SU(2)_{L}$ singlet superfield, $\chi$. Here, the modular symmetry plays a crucial role as the Yukawa couplings acquire modular forms, which are expressed in terms of Dedekind eta function $\eta(\tau)$. Therefore, the Yukawa couplings follow transformations akin to other matter fields, thereby obviating the necessity of additional flavon fields. The acquisition of $vev$ by complex modulus $\tau$ leads to the breaking of $A_{4}$ modular symmetry. We have obtained predictions on neutrino oscillation parameters, for example, the normal hierarchy for the neutrino mass spectrum. Furthermore, we find that heavy Dirac fermions, in our model, can decay to produce observed baryon asymmetry of the Universe through Dirac leptogenesis.

\end{abstract}
\section{Introduction} In the last two decades several neutrino experiments have confirmed the existence of large leptonic mixing and non-zero neutrino mass. However, there are still three significant questions about neutrino physics that remain unresolved. These questions are (a) neutrino mass hierarchy - whether it's normal hierarchy ($m_1<m_2<m_3$) or inverted hierarchy ($m_3<m_1<m_2$), and (b) the $CP$-violation in the leptonic sector, (c) octant degeneracy of atmospheric mixing angle ($\theta_{23}$). The latest global fit analysis discussed in Ref. \cite{Esteban:2020cvm} provides information about the experimental status of various neutrino oscillation observables. Another lingering question pertains to the nature of neutrinos $i.e.$ whether they are of Dirac or Majorana type? While the Majorana nature holds theoretical favour, empirical observations to date align with the Dirac nature of neutrinos \cite{Dolinski:2019nrj}. The experimental investigations, particularly those concerning neutrinoless double beta ($0\nu\beta\beta$) decay, have yielded null results, which could otherwise confirm the Majorana nature of neutrinos \cite{Dolinski:2019nrj}.\\
While the absence of positive outcomes in experiments involving $0\nu\beta\beta$ decay does not definitively establish the Dirac nature of light neutrinos, it does provide significant hints. These hints are substantial enough to formulate scenarios that predict Dirac neutrinos possessing the appropriate mass and mixing properties. Several proposals have been suggested, utilizing additional symmetries such as $U(1)_\text{{B-L}}$, $A_{4}$, and $Z_{N}$, aimed at producing small Dirac neutrino masses \cite{Roncadelli:1983ty,Davidson:2009ha,Chen:2012jg,Memenga:2013vc,Ma:2014qra,Valle:2016kyz,CentellesChulia:2016rms,Reig:2016ewy,CentellesChulia:2017koy,CentellesChulia:2016fxr, Borah:2017leo,Bonilla:2017ekt,Borah:2017dmk,Borah:2018nvu, Borah:2019bdi,CentellesChulia:2022vpz}. These symmetries play a critical role in preventing the existence of a Majorana mass term ($\Bar{\nu}_{R}^{c}\nu_{R}$) among right-handed neutrinos, as well as a tree-level Dirac mass term ($Y_{\nu}\Bar{L}\Tilde{\Phi}\nu_{R}$) between left-handed $SU(2)_{L}$ doublet and right-handed neutrino field. This prevention is necessary because achieving such a small neutrino mass $m_{\nu} \lesssim\mathcal{O}(0.1)$ eV would require Yukawa couplings $Y_{\nu} \lesssim\mathcal{O}(10^{-12})$, which is unnatural. For a detailed review of neutrino mass models based on non-abelian discrete symmetries see \cite{Ishimori:2010au,King:2015aea,deGouvea:2016qpx,Chauhan:2023faf} and references therein. An alternative approach based on modular symmetry was introduced in Ref. \cite{Feruglio:2017spp} as a generalization of discrete symmetry group approach to explain mass and mixing structure of lepton flavors. The later approach has the drawback of being dependent on adhoc symmetry breaking patterns and vacuum alignments acquired by the additional flavon fields present in the scalar sector of the theory. In the former setting, on the other hand, flavor symmetry is realized through nonlinear transformations with minimal and well-defined symmetry breaking pattern. In this approach, all Yukawa couplings are treated as modular forms with even weights, depending on complex modulus $\tau$, which describes the geometry of compactified
extra-dimensions in superstring theory. The Lagrangian maintains invariance under the finite modular symmetry group $\Gamma_{N}$. This group, corresponding to a specific value of $N$, is isomorphic to non-abelian discrete symmetry groups, such as $\Gamma_{2}\simeq S_{3}$\cite{Kobayashi:2018vbk,Kobayashi:2018wkl,Okada:2019xqk,Kobayashi:2019rzp,Marciano:2020qjf,Behera:2024ark}, $\Gamma_{3}\simeq A_{4}$ \cite{Kobayashi:2018scp,Novichkov:2018yse,Nomura:2019jxj,Ding:2019zxk,Ding:2019gof,Kobayashi:2019xvz,Nomura:2019lnr,Asaka:2020tmo,Okada:2019uoy,Nomura:2019yft,Mishra:2022egy,Behera:2020lpd,Kashav:2022kpk,Kumar:2023moh,Devi:2023vpe,CentellesChulia:2023osj}, $\Gamma_{4}\simeq S_{4}$\cite{Penedo:2018nmg,Novichkov:2018ovf,deMedeirosVarzielas:2019cyj,Kobayashi:2019mna,King:2019vhv,Criado:2019tzk,Wang:2020dbp}, and $\Gamma_{5}\simeq A_{5}$\cite{Novichkov:2018nkm,Ding:2019xna}. However, the overall symmetry will be broken when the modulus $\tau$ acquires its vacuum expectation value ($vev$) rather than the flavons (in non-abelian discrete symmetry approach). Interestingly, very few models exist that have been developed based on higher degree modular forms, also called Siegel Modular forms, unlike the traditional modular forms, which have degree 1. One can find these models devoid of any additional scalar field in the literature \cite{Ding:2021iqp,Ding:2020zxw,Kikuchi:2023dow,RickyDevi:2024ijc} that explore a unique approach to predict the precise value of neutrino parameters.\\
Furthermore, the Big Bang nucleosynthesis and cosmic microwave background observations suggest that the Universe contains more baryonic matter than antimatter, often known as matter-antimatter asymmetry. This asymmetry is quantified by the baryon-to-photon ratio, which is estimated to be  $\eta_B=(n_{B}-n_{\Bar{B}})/n_{\gamma}=6.1 \times 10^{-10}$\cite{Workman:2022ynf,Planck:2018vyg}. In order to generate this baryon asymmetry dynamically, any theoretical framework must adhere to a set of principles known as the Sakharov conditions \cite{Sakharov:1967dj}. However, these conditions are not satisfied in the Standard Model (SM). Leptogenesis is one beyond SM scenario that has been proposed as a possible explanation for the observed baryon asymmetry in the Universe (BAU) \cite{Fukugita:1986hr}. It operates through the generation of a lepton asymmetry, which, through the $B+L$ violating electroweak sphaleron processes leads to the production of a corresponding baryon asymmetry. Modular symmetry has been successfully employed to explain BAU assuming neutrinos to be of Majorana nature\cite{Asaka:2019vev,Kashav:2021zir,Behera:2020sfe,Wang:2019ovr,Behera:2022wco,Kang:2022psa,Ding:2022bzs}.
However, in absence of the evidences signaling Majorana nature, alternative avenues have been explored to explain successful baryogenesis with Dirac neutrinos as well\cite{Dick:1999je,Boz:2004ga,Thomas:2005rs,Gu:2006dc,Bechinger:2009qk,Chen:2011sb,Choi:2012ba,Borah:2016zbd,Gu:2016hxh,Narendra:2017uxl}.\\
In this study, we explore a Dirac mass model employing Type-I seesaw mechanism utilizing $A_{4}$ modular symmetry to bypass the need for extra flavon fields in the scalar sector and, in which Yukawa couplings are dictated by complex modulus $\tau$. The SM is extended with the inclusion of right-handed neutrinos, heavy Dirac fermions, and a scalar field, which are singlet under $SU(2)_{L}$. The fields are assigned modular weights in such a manner that both Majorana and Dirac mass term in the Lagrangian are prohibited. Our model accurately forecasts various parameters of neutrino oscillations and exclusively permits only normal hierarchy for neutrino mass ($m_1 < m_2 < m_3$).\\
 Furthermore, the absence of the lepton-number violating Majorana mass term in our model does not hinder its ability to make predictions regarding baryogenesis through leptogenesis. In our model, the heavy Dirac neutrinos can undergo $CP$-violating (due to complex modulus $\tau$) and out-of-equilibrium decay processes into both the left-handed and right-handed lepton sectors, resulting in the generation of equal and opposite asymmetry in both sectors. However, due to the singlet nature of the right-handed neutrino, only the asymmetry in the left sector contributes to baryon symmetry $via$ the electroweak sphaleron process. The mechanism is known as Dirac leptogenesis which has been widely explored in the literature using continuous and cyclic symmetries \cite{Dick:1999je,Boz:2004ga,Thomas:2005rs,Gu:2006dc,Bechinger:2009qk,Chen:2011sb,Choi:2012ba,Borah:2016zbd,Gu:2016hxh,Narendra:2017uxl}. In this work, we have investigated Dirac leptogenesis in the modular symmetry approach for the first time.\\
The paper is structured as follows: Section \ref{sec2} provides an overview of the fundamental framework of modular symmetry. In Section \ref{model}, we outline the model based on $A_4$ modular symmetry. Section \ref{Numerical} delves into numerical analysis, yielding predictions on neutrino observables. Section \ref{lepto} presents the model's projections regarding Dirac leptogenesis. In Section \ref{summary}, we summarize the main findings of the present work.

\section{Modular Symmetry}\label{sec2}
  The origin of modular symmetries may be attributed to the magnetized extra dimensions, compactification of heterotic strings and D-brane models \cite{Lauer:1989ax,Lerche:1989cs,Lauer:1990tm,Cremades:2004wa,Kobayashi:2017dyu}. The modular symmetry provides a novel framework for determining the flavor structure of masses and mixing and are, in general, non-linear realizations of flavor symmetries. The linear flavor symmetries exploit a particular vacuum alignment where symmetry remains unbroken, but in the nonlinear scenario, such a configuration fundamentally doesn't exist. In vacuum space, the flavor symmetry is broken except some residual fixed points. Each vacuum apart from fixed points holds equal potential as a candidate for describing the fermion spectrum, devoid of bias or preference toward any specific one. For fixed points, the modulus stabilization is needed so that the noval minima can have phenomenological implications \cite{Novichkov:2022wvg, King:2023snq}.

\noindent Modular symmetry represents the geometrical symmetry of two dimensional torus defined as Eucledian plane $R^{2}$ divided by the lattice $\Lambda$. If the lattice $(\Lambda)$ is spanned by basis vectors \{$e_{1}$, $e_{2}$\}, then complex modulus $\tau = e_{2}/e_{1}$ represents a torus structure. Also, $\Lambda$ can be spanned by the other set of basis {$e'_{1}$, $e'_{2}$\} such as
$$\begin{pmatrix}
e'_{2}\\
e'_{1}\\
\end{pmatrix} =\begin{pmatrix}
q & r  \\
s & t  \\
\end{pmatrix} \begin{pmatrix}
e_{2}\\
e_{1}\\
\end{pmatrix}$$ satisfying the condition $qt-rs=1$ spanning the group SL(2,Z) isomorphic to modular group ($\Gamma$). The modular group ($\Gamma$) consists of linear fractional transformations operating on the complex modulus ($\tau$) upper half-plane such as}
\begin{eqnarray}
\gamma: \tau \rightarrow \gamma(\tau)=\frac{q \tau+r}{s \tau+t}, \hspace{0.5cm} (qt-rs=1, \hspace{0.3cm} \text{with $q,r,s,t$}\hspace{0.3cm} \text{as integers}).
\end{eqnarray}
The complex number $\tau$ serves as a complex modulus with $-0.5\leq \text{Re}[\tau]\leq0.5$, $\text{Im}[\tau]>0$ and $|\tau| \geq 1$ as its fundamental domain. Further, the inhomogeneous modular group $\bar{\Gamma}$ is isomorphic to projective special linear group, $\mathrm{PSL}(2, Z)=\mathrm{PSL}(2, Z)/\{I,-I\}$, consisting $2\times2$ matrices with determinant equal to 1 and $q,r,s,t$ as its elements. The modular transformation is dictated by the $S$ and $T$ transformations $S: \tau \rightarrow -\frac{1}{\tau}$ and $T: \tau \rightarrow (\tau+1)$ with $S^{2}=\mathbb{I}$ and $ST^{3}=\mathbb{I}$. The series of modular groups ($\Gamma(\mathrm{N})$) is defined as
\begin{eqnarray}
\Gamma(\mathrm{N})=\left\{\left[\begin{array}{ll}
q & r \\
s & t
\end{array}\right] \in \mathrm{SL}(2, Z), \quad\left[\begin{array}{ll}
q & r \\
s & t
\end{array}\right]=\left[\begin{array}{ll}
1 & 0 \\
0 & 1
\end{array}\right](\bmod N)\right\},
\end{eqnarray}
where $\Gamma(1)=\operatorname{SL}(2, Z)$. It is important to note that $\Gamma$ and $-\Gamma$ determine the same linear transformations. The modular group $\bar{\Gamma}(N)$ provides distinct linear transformations such that $\bar{\Gamma}=\bar{\Gamma}(1)=\Gamma(1) /\{I,-I\}=$ $\mathrm{PSL}(2, Z)$. For $N\leq 2$, it may be observed that $\bar{\Gamma}(N)=\Gamma(N) /\{I,-I\}$, and for $N>2$, $\bar{\Gamma}(N)=\Gamma(N)$. These are, also, referred as the infinite modular groups. The finite modular groups are defined as the quotient group $\Gamma_N=\bar{\Gamma} / \bar{\Gamma}(N)$. The finite modular groups $\Gamma_N$ are isomorphic to permutation groups, such that $\Gamma_2 \simeq S_3$, $\Gamma_3 \simeq A_4$, $\Gamma_4 \simeq S_4$, and $\Gamma_5 \simeq A_5$. Furthermore, meromorphic functions defined on the upper half of the complex plane preserve their characteristics under all transformations within the modular group such that $f(\gamma\tau)=f(\tau)$, where $\gamma$ denotes a transformation. Due to restrictive nature of such transformation, we ask for meromorphic functions $f(\tau)$ so that after transformations, $f(\gamma\tau)$ retains the same zeros and poles as $f(\tau)$. Modular forms $f(\tau)$ are holomorphic functions with weight $2k$ and level $N$, possessing well-defined transformation properties under the group $\Gamma(N)$.  For level $N = 3$ and $k\geq 0$,
\begin{equation}
f(\gamma \tau)= (s\tau + t)^{2k} f(\tau).
\end{equation}
such that modular forms remain invariant under the infinite modular group $\Gamma(3)$ up to a factor of $(s \tau + t)^{2k}$, but they do transform under $\Gamma_{3}$. Each invariant term should have the vanishing sum of modular weight. With $N=3$, $\Gamma_{3}\simeq A_{4}$ serves as a non-linear realization of A$_4$ non-Abelian discrete symmetry. The linear space of modular forms having modular weight $2k$ and level $k$ has dimension $2k+1$. This results in modular forms for different weights as given in Appendix \ref{appdx1}.

\section{The Model}\label{model}
The field assignments within Type-I Dirac seesaw model, subject to the $A_{4}$ modular symmetry in the supersymmetric context, are given in Table \ref{tab1}. The fermionic sector of the SM is extended by three right-handed neutrinos ($\nu^{c}_{iR}$) and three heavy Dirac fermions ($N_{iL}, N^{c}_{iR}$) superfields, all are singlet under $SU(2)_{L}$ symmetry. The scalar sector is extended by the inclusion of a $SU(2)_{L}$ singlet superfield  $\chi$. We chose charge assignments to the superfields within modular symmetry to disallow the Yukawa term $L\Phi_{u}\nu^{c}_{R}$ and prevent the emergence of Majorana mass term for neutrinos, thereby ensuring Dirac nature of the model. The tensor products of the $A_{4}$ group are given in Appendix \ref{appdx2}. The Feynman diagram representing the Type-I Dirac seesaw is shown in Fig. \ref{diractree}.
\begin{table}[]
    \centering
    \begin{tabular}{c| c c c c c c| c c c| c c}
    \hline
    \hline
    Symmetry & $L_{e}$ &$L_{\mu}$ & $L_{\tau}$ & $e_{R}^{c}$ & $\mu_{R}^{c}$ & $\tau_{R}^{c}$ & $\nu^{c}_{i R}$ & $N_{i L}$ & $N^{c}_{i R}$ & $\Phi_{u,d}$ & $\chi$ \\
     \hline
    $\mathrm{A}_4$ & 1 & $1''$ & $1'$ & 1 & $1'$ & $1''$ & 1, $1'$, $1''$ & 1, $1''$, $1'$ & 1, $1'$, $1''$ & 1 & 1\\
    \hline
    $\kappa_{I}$ & -5 & -5 & $-5$ & -1 & -1 & -1 & 2 & 1 & -1 & 0 & -7\\
    \hline
    \hline
    \end{tabular}
    \caption{Field content and charge assignments of the Type-I Dirac seesaw model under $A_{4}$ and $\kappa_{I}$.}
    \label{tab1}
\end{table}

\begin{table}
\centering
\begin{tabular}{c|cc|c}
\hline\hline
$\mathbf{Y_a^b}$ & $\mathbf{Y_1^4}$ & $\mathbf{Y_{1^{\prime}}^4}$ & $\mathbf{Y_1^6}$\\
\hline $\mathrm{A}_4$ & 1 & $1^{\prime}$ & 1  \\
\hline$-k_I$ & 4 & 4 & 6 \\
\hline\hline
\end{tabular}
\caption{The higher-order Yukawa couplings transformations within the $A_4$ modular symmetry.}
\label{tab2}
\end{table}
\noindent The superpotential for the charged leptons is given by
\begin{eqnarray}
\mathcal{W}_{l} = \alpha(\mathbf{Y_{1}^{6}}L_{e}\Phi_{d} e_{R}^{c})+\beta(\mathbf{Y_{1}^{6}}L_{\mu}\Phi_{d} \mu_{R}^{c})+\gamma(\mathbf{Y_{1}^{6}}L_{\tau}\Phi_{d} \tau_{R}^{c})
\end{eqnarray} 
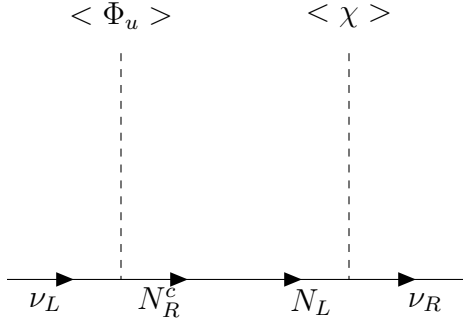
\begin{figure}[t]
    \centering
   \begin{tikzpicture}
\begin{feynman}
\vertex at (0,0) (a);
\vertex at (1.5,0) (i1);
\vertex at (-1.5,0) (i2);
\vertex at (1.5,0) (c1);
\vertex at (-1.5,0) (c2);
\vertex at (3,0) (d);
\vertex at (-3,0) (e);
\vertex at (1.5,3) (f1);
\vertex at (-1.5,3) (f2);
\vertex at (1.5,3.5) () {\(<\chi>\)};
\vertex at (-1.5,3.5) () {\(<\Phi_{u}>\)};
\vertex at (2.5,-0.3) () {\(\nu_{R}\)};
\vertex at (-2.5,-0.3) () {\(\nu_{L}\)};
\vertex at (1,-0.3) () {\(N_{L}\)};
\vertex at (-1,-0.3) () {\(N^{c}_{R}\)};
\diagram*{
(a) -- [fermion] (i1), (i2) -- [fermion] (a),
(c1) -- [fermion] (d), (e) -- [fermion] (c2),  (f1) -- [scalar] (c1), (f2) -- [scalar] (c2),
};
\end{feynman}
\end{tikzpicture}
    \caption{Tree-level Feynman diagram representing generation of neutrino mass in Type-I Dirac seesaw model.}
    \label{diractree}
\end{figure}
and the corresponding charged lepton mass matrix is given by
\begin{eqnarray}
M_{l} =\frac{v_{d} \mathbf{Y_{1}^{6}}}{\sqrt{2}} 
\begin{pmatrix}
\alpha & 0  & 0\\
0 & \beta & 0 \\
0 & 0  & \gamma \\
\end{pmatrix},
\end{eqnarray}
where $v_{d}$ and $v_{u}$ are the vacuum expectation values ($vev$) of Higgs superfields $\Phi_{d}$ and $\Phi_{u}$, respectively (\textit{i.e.} $<\Phi_{(u,d)}>=\frac{v_{(u,d)}}{\sqrt{2}}$), and assuming $v_{u}=v_{d}$ with $v_{\phi}=\sqrt{v_{u}^{2}+v_{d}^{2}}=174$ GeV and $\mathbf{Y_{1}^{6}}$ is the Yukawa coupling transforming as singlet under $A_{4}$ symmetry with modular weight 6.
Also, the Yukawa Lagrangian of the neutrino sector is given by
\begin{eqnarray}
\nonumber
\mathcal{W}_{\nu}&=&h_1\left(\mathbf{Y_{1}^{6}}L_{e}\Phi_{u}N^{c}_{1R}\right)+h_2\left(\mathbf{Y_{1}^{6}}L_{\mu}\Phi_{u}N^{c}_{2R}\right)+h_3\left(\mathbf{Y_{1}^{6}}L_{\tau}\Phi_{u}N^{c}_{3R}\right)\\
\nonumber
&+&h_{11}\left(\mathbf{Y_{1}^{4}}N_{1L}\chi \nu^{c}_{1R}\right)+ h_{13}\left(\mathbf{\mathbf{Y_{1'}^{4}}}N_{1L}\chi \nu^{c}_{3R}\right)+h_{21}\left(\mathbf{Y_{1'}^{4}}N_{2L}\chi \nu^{c}_{1R}\right)\\ \nonumber
&+&h_{22}\left(\mathbf{Y_{1}^{4}}N_{2L}\chi \nu^{c}_{2R}\right)+h_{32}\left(\mathbf{Y_{1'}^{4}}N_{3L}\chi \nu^{c}_{2R}\right)+h_{33}\left(\mathbf{Y_{1}^{4}}N_{3L}\chi \nu^{c}_{3R}\right)\\
&+&M_{1}N_{1L}N^{c}_{1R}+M_{2}N_{2L}N^{c}_{2R}+M_{3}N_{3L}N^{c}_{3R}.
\end{eqnarray}
The corresponding Dirac neutrino mass matrices can be written
\begin{eqnarray}\label{eqn7}
M_{D} =v_{u}\hspace{0.1cm}Y_{L}=v_{u} \mathbf{Y_{1}^{6}}
\begin{pmatrix}
h_1  & 0  & 0\\
0 & h_2 & 0 \\
0 & 0  & h_3 \\
\end{pmatrix},\hspace{1cm} \text{and} \hspace{1cm}
M_{D'} =u \hspace{0.1cm}Y_{R}=u
\begin{pmatrix}
h_{11} \mathbf{Y_{1}^{4}} & 0  & h_{13} \mathbf{Y_{1'}^{4}}\\
h_{21} \mathbf{Y_{1'}^{4}} & h_{22} \mathbf{Y_{1}^{4}} & 0 \\
0 & h_{32} \mathbf{Y_{1'}^{4}}  & h_{33} \mathbf{Y_{1}^{4}}\\
\end{pmatrix},
\label{md}
\end{eqnarray}
and the mass matrix for the heavy Dirac fermions is given by
\begin{eqnarray}\label{eqn88}
M =
\begin{pmatrix}
M_1 & 0  & 0\\
0 & M_2 & 0 \\
0 & 0  & M_3\\
\end{pmatrix},
\end{eqnarray}
The transformations of the higher-order Yukawa couplings within $A_4$ modular symmetry are given in Table \ref{tab2}.

\noindent In Eqns. (\ref{eqn7}) and (\ref{eqn88}), the parameters $h_1$, $h_2$, and $h_3$ can be absorbed into $h'_{ij}$ by defining: $h'_{i1} = h_{i1} h_1$, $h'_{i2} = h_{i2} h_2$, and $h'_{i3} = h_{i3} h_3$. The neutrino mass matrix can be obtained from the Type-I Dirac seesaw mechanism \cite{CentellesChulia:2017koy} given by
\begin{eqnarray}\label{eqn9}
m_{\nu'}&=-M_{D'}M^{-1}M_{D},\\
&=
\begin{pmatrix}
a & b & c \\
 d & e & f \\
 g & h & j \\
\end{pmatrix},
\end{eqnarray}
where elements of $m_{\nu'}$ are given by
\begin{equation}\label{eqn11}
\begin{rcases}
    a&=\frac{{h'_{11}}^2 u^2 v_{u}^2 \left({\mathbf{Y_{1}^{4}}}\right)^{2} \left({\mathbf{Y_{1}^{6}}}\right)^{2}}{M_{1}^{2}}+\frac{{h'_{13}}^2 u^2 v_{u}^2 \left({\mathbf{Y_{1'}^{4}}}\right)^{2} \left({\mathbf{Y_{1}^{6}}}\right)^{2}}{M_{3}^{2}}, \\
    b&=\frac{h'_{11} h'_{21} u^2 v_{u}^2 \mathbf{Y_{1}^{4}} \left({\mathbf{Y_{1}^{6}}}\right)^{2} {\mathbf{Y_{1'}^{4}}}}{M_{1}^{2}}, \\
    c&=\frac{h'_{13} h'_{33} u^2 v_{u}^2 \mathbf{Y_{1}^{4}} \left({\mathbf{Y_{1}^{6}}}\right)^{2} {\mathbf{Y_{1'}^{4}}}}{M_{3}^{2}},\\
    d&=\frac{h'_{11} h'_{21} u^2 v_{u}^2 \mathbf{Y_{1}^{4}} \left({\mathbf{Y_{1}^{6}}}\right)^{2} {\mathbf{Y_{1'}^{4}}}}{M_{1}^{2}},\\
    e&=\frac{{h_{21}'}^2 u^2 v_{u}^2 \left({\mathbf{Y_{1'}^{4}}}\right)^{2} \left({\mathbf{Y_{1}^{6}}}\right)^{2}}{M_{1}^{2}}+\frac{{h_{22}'}^2 u^2 v_{u}^2 \left({\mathbf{Y_{1}^{4}}}\right)^{2} \left({\mathbf{Y_{1}^{6}}}\right)^{2}}{M_{2}^{2}},\\ 
    f&=\frac{h'_{22} h'_{32} u^2 v_{u}^2 \mathbf{Y_{1}^{4}} \left({\mathbf{Y_{1}^{6}}}\right)^{2} {\mathbf{Y_{1'}^{4}}}}{M_{2}^{2}},\\
    g&=\frac{h'_{13} h'_{33} u^2 v_{u}^2 \mathbf{Y_{1}^{4}} \left({\mathbf{Y_{1}^{6}}}\right)^{2} {\mathbf{Y_{1'}^{4}}}}{M_{3}^{2}},\\
    h&=\frac{h'_{22} h'_{32} u^2 v_{u}^2 \mathbf{Y_{1}^{4}} \left({\mathbf{Y_{1}^{6}}}\right)^{2} {\mathbf{Y_{1'}^{4}}}}{M_{2}^{2}},\\ 
    j&=\frac{{h_{32}'}^2 u^2 v_{u}^2 \left({\mathbf{Y_{1'}^{4}}}\right)^{2} \left({\mathbf{Y_{1}^{6}}}\right)^{2}}{M_{2}^{2}}+\frac{{h_{33}'}^2 u^2 v_{u}^2 \left({\mathbf{Y_{1}^{4}}}\right)^{2} \left({\mathbf{Y_{1}^{6}}}\right)^{2}}{M_{3}^{2}}.
\end{rcases}
\end{equation}
\noindent We define the Hermitian neutrino mass matrix as
\begin{eqnarray}
m_{\nu}\equiv m_{\nu'}m_{\nu'}^{\dagger}=\left(
\begin{array}{ccc}
 a^2+b^2+c^2 & a d+b e+c f & a g+b h+c j \\
 a d+b e+c f & d^2+e^2+f^2 & d g+e h+f j \\
 a g+b h+c j & d g+e h+f j & g^2+h^2+j^2 \\
\end{array}
\right).
\label{mnu}
\end{eqnarray}
In the following section, we perform numerical analysis and explore the model's prediction for various neutrino oscillation observables utilizing the diagonalization constraints of neutrino mass matrix given in Eqn. (\ref{mnu}).

\section{Numerical Analysis and Discussion}\label{Numerical}
In numerical analysis, the real coupling constants, which are free parameters in the model, are randomly varied within the range:

\begin{eqnarray}
h'_{ij}\in (-1, 1)  \hspace{0.5cm} \text{and} \hspace{0.5cm} \alpha,\beta,\gamma\in (0, 1).
\end{eqnarray}
\noindent In addition, the Yukawa couplings entering in $Y_{L}$ and $Y_{R}$ (in Eqn. (\ref{md})) depend on complex modulus $\tau$. The real and imaginary components of $\tau$ are randomly varied within the fundamental domain, $-0.5\leq \text{Re}[\tau]\leq0.5$ and $\text{Im}[\tau]>0$. The $vev$ of singlet scalar field $\chi$ is varied randomly in the range ($1-10^{6}$) GeV. In order to determine the masses of the active light neutrinos ($m_{1}, m_{2},m_{3}$), we have performed numerical diagonalization of the neutrino mass matrix ($m_{\nu}$) given in Eqn. (\ref{mnu}). The Hermitian neutrino mass matrix $m_{\nu}$ can be, in general, diagonalized by unitary mixing matrix $U$,\textit{viz.}, $Um_{\nu}U^{\dagger} = \text{diag}(m_{1}^{2}, m_{2}^{2}, m_{3}^{2})$. The neutrino mixing matrix $U$ can be parameterized in terms of three mixing angles ($\theta_{12}$, $\theta_{13}$, $\theta_{23}$) and one Dirac-type $CP$-violating phase ($\delta$) in the charged lepton basis
\begin{equation}
U =
\begin{pmatrix}
c_{12}c_{13} & s_{12}c_{13}  & s_{13}e^{-i\delta} \\
-s_{12}c_{23}-c_{12}s_{23}s_{13}e^{i\delta} & c_{12}c_{23}-s_{12}s_{23}s_{13}e^{i\delta}  & s_{23}c_{13} \\
s_{12}s_{23}-c_{12}c_{23}s_{13}e^{i\delta} & -c_{12}s_{23}-s_{12}c_{23}s_{13}e^{i\delta}  & c_{23}c_{13} 
\end{pmatrix}=\begin{pmatrix}
U_{11} & U_{12}  & U_{13} \\
U_{21} & U_{22}  & U_{23} \\
U_{31} & U_{32}  & U_{33}\\
\end{pmatrix} ,
 \label{2}
\end{equation}\\
where $c_{ij}=\cos{\theta_{ij}}$ and $s_{ij}=\sin{\theta_{ij}}$ and $\delta$ is Dirac-type $CP$-violating phase.
Further, the neutrino mixing angles can be determined from the unitary matrix $U$ using the following relations
\begin{eqnarray}\label{angle}
\sin^{2}{\theta_{13}}=\left|U_{13}\right|^{2}, \hspace{0.5cm} \sin^{2}{\theta_{23}}=\dfrac{\left|U_{23}\right|^{2}}{1-\left|U_{13}\right|^{2}}, \hspace{0.5cm} \sin^{2}{\theta_{12}}=\dfrac{\left|U_{12}\right|^{2}}{1-\left|U_{13}\right|^{2}}.
\label{eqn16}
\end{eqnarray}
Also, the degree of $CP$-violation quantified using Jarlskog invariant \cite{Krastev:1988yu,Jarlskog:1985ht} is defined as
\begin{equation}
J_{CP} = \text{Im}\left[U_{11}U_{22}U^{*}_{12}U^{*}_{21}\right] = s_{12}c_{12}s_{23}c_{23}s_{13}c^{2}_{13}\sin{\delta}.
\label{eqn17}
\end{equation}
\begin{figure}[t]
    \centering
    \includegraphics[height=5cm,width=7cm]{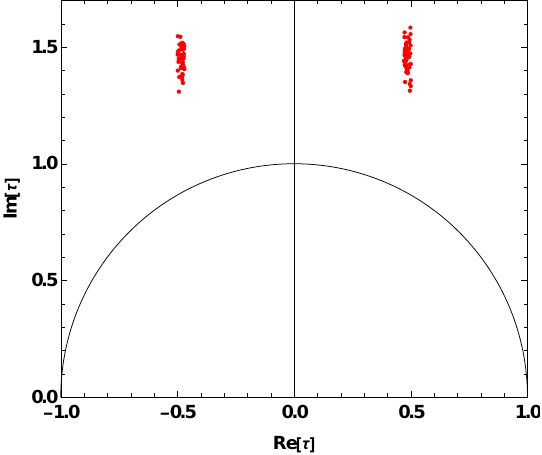}
    \caption{The parameter space of real and imaginary parts of complex modulus $\tau$ within the fundamental domain.}
    \label{tau}
\end{figure}
\begin{figure}[t]
    \centering
\includegraphics[scale=0.8]{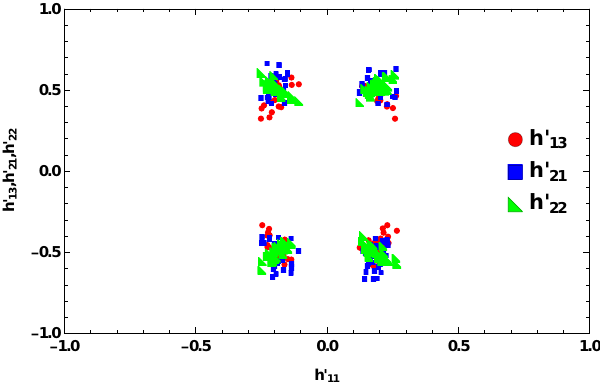}
\includegraphics[scale=0.8]{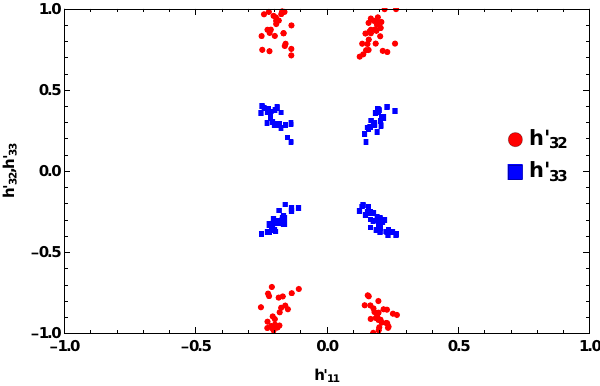}
    \caption{Allowed ranges of parameters $h'_{ij}$ from the neutrino oscillation data shown in Table \ref{tabx}.}
    \label{h}
\end{figure}
\begin{table}[t]
    \centering
    \begin{tabular}{l l l l l}
    \hline
    \hline
  Parameter &best-fit$\pm 1\sigma$ range (NH)&best-fit$\pm 1\sigma$ range (IH)& $3\sigma$ range (NH) & $3\sigma$ range (IH)\\
\hline  
\hline
 $\sin^{2}{\theta_{12}}$&$0.304^{+0.013}_{-0.012}$&$0.304^{+0.012}_{-0.012}$ & 0.269-0.343 & 0.269-0.343 \\
\hline
$\sin^{2}\theta_{23}$&$0.573^{+0.018}_{-0.023}$&$0.578^{+0.017}_{-0.021}$& 0.405-0.624& 0.410-0.623\\

\hline
$\sin^{2}\theta_{13}$&$0.02220^{+0.00068}_{-0.00062}$&$0.02238^{+0.00064}_{-0.00062}$& 0.02060-0.02435& 0.02053-0.02434\\
\hline

$\frac{\Delta m_{23}^{2}}{10^{-3}\text{eV}^{2}}$&$2.515^{+0.028}_{-0.028}$&$2.498^{+0.028}_{-0.029}$& 2.431-2.598 &-2.584- -2.413\\

\hline
$\frac{\Delta m_{12}^{2}}{10^{-5}\text{eV}^{2}}$ & $7.42^{+0.21}_{-0.20}$&$7.42^{+0.21}_{-0.20}$& 6.82-8.04  & 6.62-8.04\\
\hline
\hline
\end{tabular}
    \caption{The neutrino oscillation data from global fit is used in the numerical analysis \cite{Esteban:2020cvm}.}
    \label{tabx}
\end{table}

\begin{figure}[t]
    \centering
\includegraphics[scale=0.85]{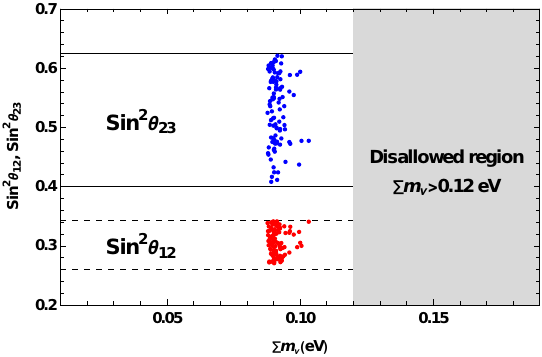}
\includegraphics[scale=0.85]{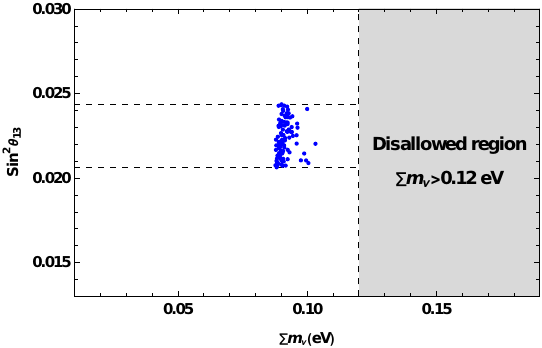}
    \caption{\textbf{Left:} Correlation plot between ($\sin^{2}{\theta_{12}},\sin^{2}{\theta_{23}}$-$\sum m_{\nu}$), \textbf{Right:} Correlation plot between ($\sin^{2}{\theta_{13}}$-$\sum m_{\nu}$), for NH.}
    \label{angles}
\end{figure}
\begin{figure}[t]
    \centering
    \includegraphics[height=5cm,width=7cm]{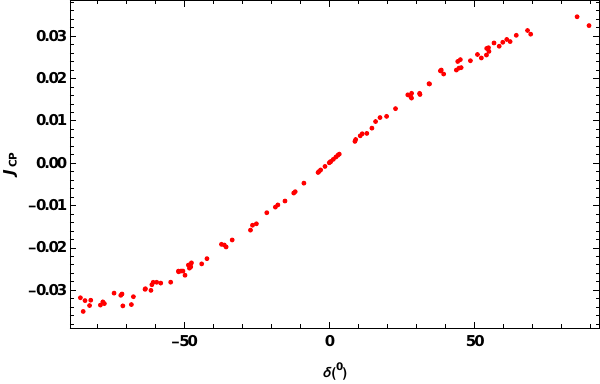}
    \caption{Correlation plot between ($\delta-J_{CP}$).}
    \label{delta}
\end{figure}
\begin{figure}[t]
    \centering
\includegraphics[scale=0.85]{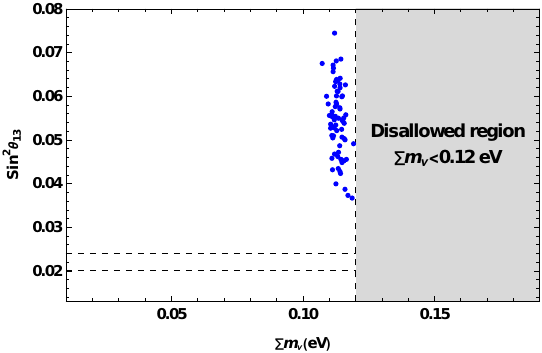}
\includegraphics[scale=0.85]{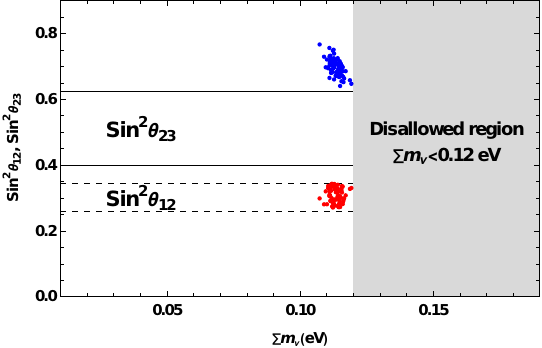}
    \caption{\textbf{Left:} Correlation plot between ($\sin^{2}{\theta_{13}}$-$\sum m_{\nu}$), \textbf{Right:} Correlation plot between ($\sin^{2}{\theta_{12}},\sin^{2}{\theta_{23}}$-$\sum m_{\nu}$), for IH.}
    \label{IH}
\end{figure}
\noindent This process yielded predictions on the neutrino mixing angles and $CP$-violation ($\delta$) using Eqns. (\ref{angle}) and (\ref{eqn17}), respectively. In Fig. \ref{tau}, we have shown the allowed parameter space of $\tau$ in the complex plane. It can be seen from Fig. \ref{tau} that Im[$\tau$], which is responsible for $CP$-violation in the model, assumes values slightly greater than 1 while the Re[$\tau$] has values near $\pm0.5$. Also, Fig. \ref{h} depicts the allowed ranges of the free parameters $h'_{ij}$ constrained by neutrino oscillation data given in Table \ref{tabx}. Further, Fig. \ref{angles} illustrates the model's prediction regarding the neutrino mixing angles. It is apparent from Fig. \ref{angles} that all the mixing angles fall within the \(3\sigma\) experimental range given in Table \ref{tabx} for NH. The value of the sum of neutrino masses lies in the range $0.088 \text{ eV}\leq\sum m_{\nu}\leq0.105 \text{ eV}$ which is below the cosmological bound on the sum of neutrino masses $\sum m_{\nu}<0.12$ eV\cite{Planck:2018vyg}. Fig. \ref{delta} illustrates the correlation between \( \delta \) and \( J_{CP} \). The $CP$-violating phase \( \delta \) is constrained to the first and fourth quadrants with the normal hierarchy (NH) only, consistent with the T2K experiment (prefer NH) at \( 3\sigma \) with nearly maximal \( CP \)-violations \cite{T2K:2019bcf, T2K:2021xwb, T2K:2023smv}. We, also, explored the parameter space pertaining to the inverted hierarchy (IH) of neutrino masses. The model ruled out the IH at $3\sigma$ confidence level as evident in Fig. \ref{IH}. The atmospheric mixing angle ($\sin^2{\theta_{23}}$) and reactor mixing angle ($\sin^2{\theta_{13}}$) falls outside $3\sigma$ experimental range (Fig. \ref{IH}), though the values of the solar mixing angle ($\sin^2{\theta_{12}}$) are consistent with $3\sigma$ experimental ranges given in Table \ref{tabx}.

\section{Dirac Leptogenesis}\label{lepto}
In this model, the lepton number conservation ensures the absence of any net lepton asymmetry. However, it is feasible to induce an equal and opposite lepton asymmetry within both left and right-handed lepton sectors ($\eta_{\Delta L}=-\eta_{\Delta \nu_{R}}$) through out-of-equilibrium and $CP$-violating decays ($N \rightarrow L \Phi$ and $N\rightarrow \nu_R \chi$) involving the heavy Dirac fermions $N_{i}$. Subsequently, the electroweak sphaleron processes \cite{Kuzmin:1985mm} facilitate the conversion of the asymmetry in the left-lepton sector into the baryon asymmetry before the electroweak phase transition. Conversely, the asymmetry in the right-lepton sector cannot be converted into baryon asymmetry due to the singlet nature of the right-handed neutrinos. We have considered the scenario wherein supersymmetry is broken at a higher mass scale exceeding the mass scale of heavy Dirac fermions ($N_{i}$) in our model. This ensures that we need not account for the contribution of superparticles to the final baryon asymmetry \cite{Giudice:2003jh,Marciano:2024quu}. In the analysis, we have considered scale of $N_1$ to be $\mathcal{O}(10^{12})$ GeV, to have leptogenesis in the unflavored regime \cite{Abada:2006fw,Nardi:2006fx,Abada:2006ea,Blanchet:2006be}. Fig. \ref{dc} illustrates the relevant decay channels of the $N_{i}$.
The $CP$ asymmetry parameter for the right-sector is given as \cite{Cerdeno:2006ha}
\begin{eqnarray}
\epsilon=\frac{\sum_{k}\left(\Gamma(N_{i}\rightarrow\nu_{Rk}\chi)-\Gamma(\bar{N_{i}}\rightarrow\bar{\nu}_{Rk}\chi^*)\right)}{\sum_{j}\Gamma(N_{i}\rightarrow\nu_{Rj}\chi)+\sum_{l}\Gamma(N_{i}\rightarrow L_{l}\Phi)}    
\end{eqnarray}
which can be simplified as 
\begin{eqnarray}
\epsilon \simeq -\frac{1}{8\pi}\frac{M_{1}}{v_{u}u}\frac{\text{Im}\left[(Y_{L}m_{\nu}^{\dagger}Y_{R})_{11}\right]}{(Y_{L}Y_{L}^{\dagger})_{11}+(Y_{R}^{\dagger}Y_{R})_{11}}.
\end{eqnarray}
The couplings $Y_L$ and $Y_R$ are functions of complex modulus $\tau$ \textit{via} modular forms $\mathbf{Y_1^6}$ and ($\mathbf{Y_1^4,Y_{1'}^4}$), respectively (refer to Appendix \ref{appdx1}).
It is possible to create a net baryon asymmetry if net lepton asymmetry is generated prior to the sphaleron decoupling epoch. However, in order to impede the left-sector lepton asymmetry from being washed out, it is important to prevent the equilibration of left and right-sectors, leading to condition
\begin{eqnarray}
\Gamma_{L-R} \sim \frac{|Y_L|^{2}|Y_R|^{2}T^{3}}{M_{1}^{2}}<H(T),\hspace{0.5cm} \text{where}\hspace{0.5cm} H=1.67\frac{\sqrt{g_*}}{M_{pl}}T^{2} \hspace{0.2cm} \text{is Hubble constant},
\end{eqnarray}
and $g_{*}$ is the effective number of relativistic degree of freedom and $M_{P}$ is the Planck mass, $M_{P}=1.2\times 10^{19}$ GeV.  

\begin{figure}[t]
    \centering
   \begin{tikzpicture}
\begin{feynman}
\vertex at (0,0) (i1);
\vertex at (-3,0) (i2);
\vertex at (2,2) (a);
\vertex at (2,-2) (b);
\diagram*{
(i2) -- [fermion, edge label=\(N_{i}\)] (i1), (i1) -- [scalar, edge label=\(\chi\)] (a), (i1) -- [fermion, edge label=\(\nu_{R}\)] (b),
};
\end{feynman}
\end{tikzpicture}
\hspace{1cm}  \begin{tikzpicture}
\begin{feynman}
\vertex at (0.5,0) (i1);
\vertex at (-1,0) (i2);
\vertex at (2,2) (a);
\vertex at (2,-2) (b);
\vertex at (-6,0) (c);
\vertex at (-3.6,0) (d);
\diagram*{
(i2) -- [fermion, edge label=\(N_{i}\)] (i1), (i1) -- [scalar, edge label=\(\chi\)] (a), (i1) -- [fermion, edge label=\(\nu_{R}\)] (b),(c) -- [fermion, edge label=\(N_{i}\)] (d),
};
\draw[decoration={markings, mark=at position 0.5 with {\arrow{<},\node[above] {L};}}, postaction={decorate}] (-1,0) arc (0:180:1.3); 

\draw[dashed, decoration={markings, mark=at position 0.5 with {\arrow{<},\node[below] {\(\Phi\)};}}, postaction={decorate}] (-1,0) arc (360:180:1.3); 
\end{feynman}
\end{tikzpicture} 
\caption{Tree-level (left) and 1-loop self-energy (right)  Feynman diagram for the decays of $N$.}
\label{dc}
\end{figure}
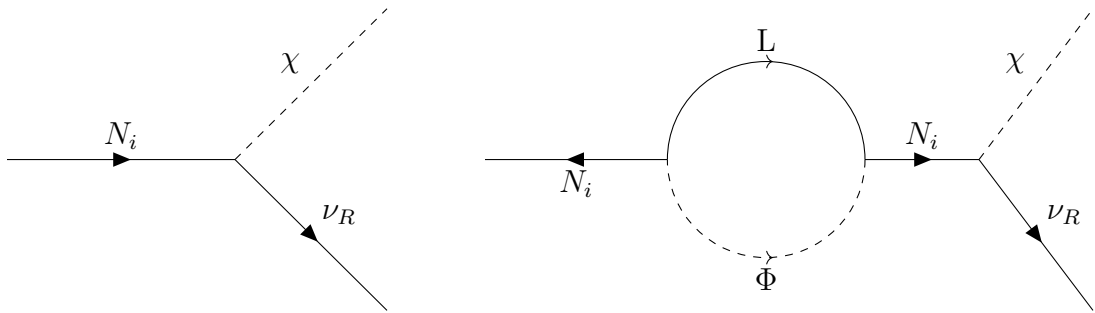
\begin{table}[t]
    \centering
    \begin{tabular}{c c c c}
    \hline
    \hline
    Benchmark Point (BP) & $M_{1}$ (GeV) &$u$ (GeV) &  Complex modulus $\tau$ \\
     \hline
     \hline
    BP1 & $10^{12}$ & $10^{3}$ & $-0.4980+1.5464i$\\
    \hline
    BP2 & $10^{12}$ & $10^{4}$ & $0.4814+1.5005i$\\
    \hline
    BP3 & $10^{12}$ & $10^{5}$ & $-0.4871+1.5123i$\\
    \hline
    \hline
    \end{tabular}
    \caption{The benchmark points utilised to produce the lepton asymmetry in the model via Dirac leptogenesis. The Yukawa couplings corresponding to BPs are given in Eqn. (\ref{yuk}).}
    \label{bp}
\end{table}
\noindent In order to realise the baryogenesis through Dirac leptogenesis in our model, we follow the scheme given in Ref.\cite{Cerdeno:2006ha}. The relevant Boltzmann equations for the Dirac leptogenesis are given by
\begin{equation}\label{bz}
\begin{rcases}
\frac{d\eta_{\Sigma N_{1}}}{dz}&=\frac{z}{H(z=1)}\left[2-\frac{\eta_{\Sigma N_{1}}}{\eta^{eq}_{N_{1}}}+\epsilon\left(\frac{3 \eta_{\Delta L}}{2}+\eta_{\Delta L}\right)\right]\Gamma^{D},\\
\frac{d\eta_{\Delta N_{1}}}{dz}&=\frac{z}{H(z=1)}\left[\eta_{\Delta L}-\frac{\eta_{\Delta N_{1}}}{\eta^{eq}_{N_{1}}}-B_{R}\left(\frac{3 \eta_{\Delta L}}{2}+\eta_{\Delta N_{1}}\right)\right]\Gamma^{D},\\
\frac{d\eta_{\Delta L}}{dz}&=\frac{z}{H(z=1)}\left\{\left[\epsilon\left(1-\frac{\eta_{\Sigma N_{1}}}{2\eta^{eq}_{N_{1}}}\right)-\left(1-\frac{B_{R}}{2}\right)\left(\eta_{\Delta L}-\frac{\eta_{\Delta N_{1}}}{\eta^{eq}_{N_{1}}}\right)\right]\Gamma^{D}-\left(\frac{3 \eta_{\Delta L}}{2}+\eta_{\Delta N_{1}}\right)\Gamma^{W} \right\}
\end{rcases}
\end{equation}
where $\eta_{\Sigma N}=\frac{\eta_{N}+\eta_{\bar{N}}}{\eta_{\gamma}}$, $\eta_{\Delta N}=\frac{\eta_{N}-\eta_{\bar{N}}}{\eta_{\gamma}}$, $z=\frac{M_{N_{1}}}{T}$ and
\begin{eqnarray}
\Gamma^{D}=\frac{z^{2}}{2}K_{1}(z)\left[\Gamma(N_{1}\rightarrow L\Phi)+\Gamma(N_{1}\rightarrow\nu_{R}\chi)\right]=\Gamma^{W}.
\end{eqnarray}
It is to be noted that the Boltzmann equations given in Eqn. (\ref{bz}) are, in general, function of complex modulus $\tau$ through $\epsilon$, branching ratio ($B_R$) and $\Gamma_D$ or $\Gamma_W$.
The electroweak sphaleron processes convert the net lepton asymmetry in the left-sector to the baryon asymmetry by a conversion factor given by
\begin{eqnarray}\label{bau}
\eta_{B}=-\frac{28}{79}\eta_{\Delta L}.
\end{eqnarray}
The allowed parameter space of vacuum alignment $u$ of singlet scalar $\chi$ and complex modulus $\tau$ have been obtained by requiring consistent low energy neutrino phenomenology in the model. Three such values of $u$ and $\tau$ are given as BPs of the model in Table \ref{bp}. With the evolution of the Universe, the heavy Dirac fermion begins to decay to SM particles, which leads to an increase in the comoving number density of lepton asymmetry ($\eta_{\Delta L}$). In the numerical analysis, the evolution of $\eta_{\Delta L}$ has been studied using coupled Boltzmann equations given in the Eqn. (\ref{bz}) for the three BPs (Table \ref{bp}). It can be deduced from Eqns. (\ref{eqn7}) and (\ref{eqn9}) that for smaller values of $u$ one requires larger Yukawa couplings,  or vice-versa, to satisfy neutrino phenomenology. It is to be noted that as we go from BP1 to BP3, $u$ is increasing from $10^3$ to $10^5$ GeV.

\begin{figure}[t]
    \centering
    \includegraphics[height=7cm,width=10cm]{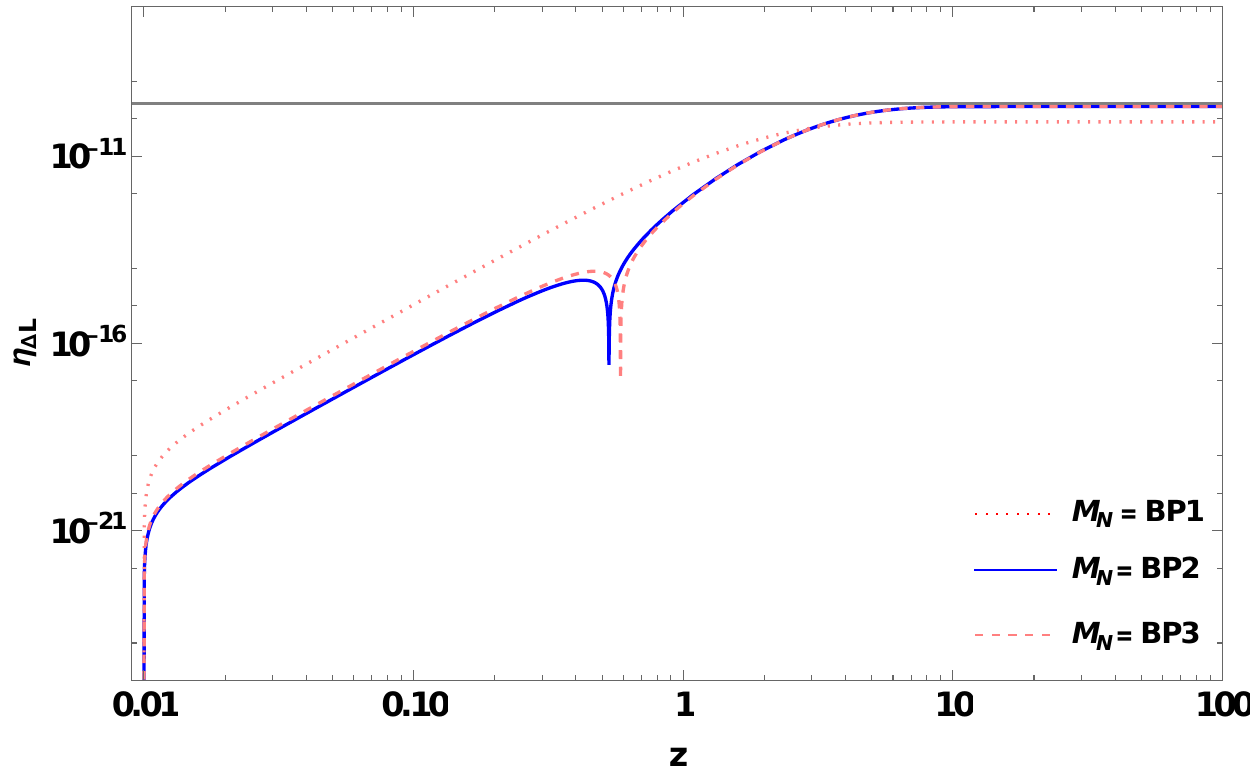}
    \caption{The evolution of comoving number density of lepton asymmetry  with $z=M_{N_{1}}/T$ for different benchmark points given in Table \ref{bp}. The required asymmetry in the lepton sector is represented by the solid horizontal line, which is then converted to the required baryon asymmetry through the electroweak sphaleron action.}
    \label{DL}
\end{figure}

\noindent In order to calculate the baryon asymmetry through Dirac leptogenesis in the model, we assume $(Y_{L}Y_{L}^{\dagger})_{11} = (Y_{R}^{\dagger}Y_{R})_{11}$. In Fig. \ref{DL}, we have depicted the evolution of the comoving number density of the lepton asymmetry as a function of $z\equiv M_{N_1}/T$. The larger Yukawa coupling (smaller $u$) while generating lepton asymmetry shall enhance washout scattering processes ($\Gamma_{W}$) as well which has been shown in Fig. (\ref{DL}) with dotted-red curve. However, as discussed earlier, the larger $u$ induces smaller Yukawa couplings which in turn diminishes washout effects culminating in the saturation of lepton asymmetry (blue and dashed-pink curves), near $z\approx 6$, consistent with observed value of $\eta_{\Delta L}$. The resulting lepton asymmetry transforms into the baryon asymmetry through the $B+L$ violating electroweak sphaleron process, as given in Eqn. (\ref{bau}).

\section{Conclusions}\label{summary}
In summary, absence of evidences, pointing towards Majorana nature of neutrinos, have motivated theoretical investigations explaining non-zero neutrino mass and BAU considering neutrino to be  Dirac particle. The baryogenesis \textit{via} leptogenesis have been widely discussed within flavor symmetries, discrete or cyclic, for recent review see \cite{deGouvea:2016qpx} and references therein. These frameworks, sometimes lacks justification for adhoc symmetry breaking patterns and vacuum alignments acquired by the additional flavon fields in the scalar sector of the model. The recent investigations, in this direction, have focused on modular symmetry wherein flavor symmetry is realized through non-linear transformations with minimal and well defined symmetry breaking pattern. In this work, we have proposed a scenario of neutrino mass generation explaining, simultaneously, the low energy neutrino phenomenology and BAU, assuming Dirac nature of neutrino. \\
In particular, we have investigated a Dirac mass model incorporating $A_{4}$ modular symmetry within the Type-I seesaw framework to elucidate neutrino phenomenology and account for observed oscillation data. The SM fermionic sector is extended by three right-handed SU(2)$_L$ singlets and three heavy Dirac fermionic superfields, while its scalar sector is augmented by a scalar superfield singlet under SU(2)$_L$ gauge symmetry. Choosing the appropriate charge assignment under $A_4$ modular symmetry, prevents the tree-level Dirac and Majorana mass terms, facilitating the construction of neutrino mass matrix within Type-I Dirac seesaw. The Yukawa couplings are related to the modular forms under $A_4$ modular symmetry. The invariance under $A_4$ modular symmetry is broken by the \textit{vev} of the complex modulus $\tau$. We have explored the allowed parameter space of the model in light of global neutrino oscillation data. The model predicts normal hierarchical neutrino mass spectrum which is in consonance with the recent observation from T2K experiment. in general, the complex modulus $\tau$ (Im($\tau$)) seeds $CP$ violation and lepton asymmetry resulting in baryon asymmetry consistent with the observed BAU. We, also, find that the Dirac-type $CP$-violating phase $\delta$ lie in the range, $(-90^o-90^o)$. The $CP$-violating and out-of-equilibrium decay of heavy Dirac fermions produces equal and opposite lepton asymmetry in both the left and right-handed lepton sectors. However, due to the singlet nature of the right-handed neutrino, asymmetry in the left-handed-lepton sector converted $via$ $B+L$ violating the electroweak sphaleron process to the baryon asymmetry of the Universe. In order to study the evolution of the lepton asymmetry, we solve the coupled Boltzmann equations for three BPs. For higher Yukawa coupling, washout effects increase, thereby limiting the lepton asymmetry below the experimentally required value. However, for smaller Yukawa couplings, washout effects get diminished and lepton asymmetry saturates around its observed value as shown in Fig. \ref{DL}.

\noindent\textbf{\Large{Acknowledgments}}
 \vspace{.3cm}\\
LS acknowledges financial support provided by Council of Scientific and Industrial Research
(CSIR) vide letter No. 09/1196(18553)/2024-EMR-I. MK would like to acknowledge Physical Research Laboratory, Ahmedabad, Govt. of India for the Post-Doctoral Fellowship vide letter No. PRL/ADMN/MK/2023.

\appendix

\section{Modular forms corresponding to $\Gamma_{3}=A_{4}$}\label{appdx1}
The Dedekind eta-function, defined in the upper complex plane, constructs modular forms of weight 2 and level 3, given as
$$
\eta(\tau)=q^{1 / 24} \prod_{n=1}^{\infty}\left(1-q^n\right)
$$
with $q=e^{i 2 \pi \tau}$, and written in terms of the $\eta(\tau)$ and its derivative as

\begin{equation}\label{23}
\begin{rcases}
 Y_1(\tau)&=\frac{i}{2 \pi}\left[\frac{\eta^{\prime}(\tau / 3)}{\eta(\tau / 3)}+\frac{\eta^{\prime}((\tau+1) / 3)}{\eta((\tau+1) / 3)}+\frac{\eta^{\prime}((\tau+2) / 3)}{\eta((\tau+2) / 3)}-\frac{27 \eta^{\prime}(3 \tau)}{\eta(3 \tau)}\right], \\
Y_2(\tau)&=-\frac{i}{\pi}\left[\frac{\eta^{\prime}(\tau / 3)}{\eta(\tau / 3)}+\omega^2 \frac{\eta^{\prime}((\tau+1) / 3)}{\eta((\tau+1) / 3)}+\omega \frac{\eta^{\prime}((\tau+2) / 3)}{\eta((\tau+2) / 3)}\right], \\ 
Y_3(\tau)&=-\frac{i}{\pi}\left[\frac{\eta^{\prime}(\tau / 3)}{\eta(\tau / 3)}+\omega \frac{\eta^{\prime}((\tau+1) / 3)}{\eta((\tau+1) / 3)}+\omega^2 \frac{\eta^{\prime}((\tau+2) / 3)}{\eta((\tau+2) / 3)}\right],
\end{rcases}
\end{equation}
where $\omega=e^{i 2 \pi / 3}$. They fulfil the condition
$$
Y_2^2+2 Y_1 Y_3=0 .
$$
These three modular forms, arranged as $\mathbf{Y_3^{2}}=\left(\begin{array}{l} Y_1 \ Y_2 \ Y_3 \end{array}\right)$ in the $\mathbf{3}$ irreducible representation of $A_4$, with indices denoting weight and multiple ($\mathbf{Y_{\text{multiplet}}^{\text{weight}}})$, facilitate construction of higher-weight modular forms using tensor products of $A_4$. For $k=4$, there are 5 linearly independent modular forms, arranged into two singlets $\mathbf{1}, \mathbf{1}^{\prime}$ and one triplet $\mathbf{3}$ of $A_4$ :
$$
\mathbf{Y_1^{4}}=Y_1^2+2 Y_2 Y_3, \quad \mathbf{Y_{1^{\prime}}^{4}}=Y_3^2+2 Y_1 Y_2, \quad \mathbf{Y_3^{4}}=\left(\begin{array}{c}
Y_1^2-Y_2 Y_3 \\
Y_3^2-Y_1 Y_2 \\
Y_2^2-Y_1 Y_3
\end{array}\right) .
$$k
For $k=6$, one has
$$
\begin{aligned}
& \mathbf{Y_1^{6}}=Y_1^3+Y_2^3+Y_3^3-3 Y_1 Y_2 Y_3, \\
& \mathbf{Y_{3,1}^{6}}=\left(Y_1^2+2 Y_2 Y_3\right)\left(\begin{array}{l}
Y_1 \\
Y_2 \\
Y_3
\end{array}\right), \quad \mathbf{Y_{3,2}^{6}}=\left(Y_3^2+2 Y_1 Y_2\right)\left(\begin{array}{l}
Y_3 \\
Y_1 \\
Y_2
\end{array}\right)
\end{aligned}
$$
and the total dimension is 7. For $k=8$, there are
$$
\begin{aligned}
& \mathbf{Y_1^{8}}=\left(Y_1^2+2 Y_2 Y_3\right)^2, \quad \mathbf{Y_{\mathbf{1}^{\prime}}^{8}}=\left(Y_1^2+2 Y_2 Y_3\right)\left(Y_3^2+2 Y_1 Y_2\right), \quad \mathbf{Y_{\mathbf{1}^{\prime \prime}}^{8}}=\left(Y_3^2+2 Y_1 Y_2\right)^2, \\
& \mathbf{Y_{\mathbf{3}, 1}^{8}}=\left(Y_1^2+2 Y_2 Y_3\right)\left(\begin{array}{c}
Y_1^2-Y_2 Y_3 \\
Y_3^2-Y_1 Y_2 \\
Y_2^2-Y_1 Y_3
\end{array}\right), \quad \mathbf{Y_{\mathbf{3}, 2}^{8}}=\left(Y_3^2+2 Y_1 Y_2\right)\left(\begin{array}{l}
Y_2^2-Y_1 Y_3 \\
Y_1^2-Y_2 Y_3 \\
Y_3^2-Y_1 Y_2
\end{array}\right),
\end{aligned}
$$
corresponding to a total dimension of 9 . For $k=10$, a total of 11 linearly independent modular forms are arranged into $A_4$ multiplets as
$$
\begin{aligned}
\mathbf{Y_1^{10}} & =\left(Y_1^2+2 Y_2 Y_3\right)\left(Y_1^3+Y_2^3+Y_3^3-3 Y_1 Y_2 Y_3\right) \\
\mathbf{Y_{1^{\prime}}^{10}} & =\left(Y_3^2+2 Y_1 Y_2\right)\left(Y_1^3+Y_2^3+Y_3^3-3 Y_1 Y_2 Y_3\right) \\
\mathbf{Y_{3,1}^{10}} & =\left(Y_1^2+2 Y_2 Y_3\right)^2\left(\begin{array}{l}
Y_1 \\
Y_2 \\
Y_3
\end{array}\right), \quad \mathbf{Y_{3,2}^{10}}=\left(Y_3^2+2 Y_1 Y_2\right)^2\left(\begin{array}{l}
Y_2 \\
Y_3 \\
Y_1
\end{array}\right) \\
\mathbf{Y_{3,3}^{10}} & =\left(Y_1^2+2 Y_2 Y_3\right)\left(Y_3^2+2 Y_1 Y_2\right)\left(\begin{array}{l}
Y_3 \\
Y_1 \\
Y_2
\end{array}\right) .
\end{aligned}
$$
\section{Tensor products of $A_{4}$ group in $T$-diagonal basis}\label{appdx2}
The multiplication rules for the representations of $A_{4}$ are as follows
\begin{eqnarray}
\nonumber
1\otimes 1'=1',\hspace{5mm} 1''\otimes 1''=1' , \hspace{5mm} 1'\otimes 1''=1, \hspace{5mm} 1''\otimes 1=1'', \hspace{5mm} 1'\otimes 1'=1''
\end{eqnarray}
\begin{eqnarray}
\nonumber
3\otimes 1'=3 , \hspace{5mm} 3\otimes 1''=3, \hspace{5mm} 3\otimes3=1\oplus 1'\oplus 1''\oplus 3_{s} \oplus 3_{a}.
\end{eqnarray}
In the $T$-diagonal basis, the Clebsch–Gordan decomposition of two triplets, $m$ = ($m_1, m_2, m_3$) and $n$ = ($n_1, n_2, n_3$) is given as
\begin{eqnarray}\label{tp}
\nonumber
&&(m\otimes n)_{1}=m_1n_1+m_2n_3+m_3n_2 ,\\
\nonumber
&&(m\otimes n)_{1'}=m_3n_3+m_1n_2+m_2n_1 ,\\
\nonumber
&&(m\otimes n)_{1''}=m_2n_2+m_1n_3+m_3n_1 ,\\
\nonumber
&&(m\otimes n)_{3_{s}}=\dfrac{1}{3}\left(2m_1n_1-m_2n_3-m_3n_2, 2m_3n_3-m_1n_2-m_2n_1, 2m_2n_2-m_1n_3-m_3n_1\right) ,\\ \nonumber
&&(m\otimes n)_{3_{a}}=\dfrac{1}{2}\left(m_2n_3-m_3n_2, m_1n_2-m_2n_1, m_1n_3-m_3n_1\right).
\end{eqnarray}

\end{document}